\documentstyle[twocolumn,aps,epsf]{revtex}
\begin{document} 
\twocolumn[\hsize\textwidth\columnwidth\hsize\csname 
@twocolumnfalse\endcsname
\title{Reciprocity Theorems for Ab Initio Force Calculations}
\author{C.Wei, Steven P.  Lewis, E.J. Mele and Andrew M.  Rappe}
\address{ Department of Physics \\
 Department of Chemistry \\ Laboratory for Research on
the Structure of Matter\\ University of Pennsylvania \\ Philadelphia, 
Pennsylvania 19104}
\maketitle
\begin{abstract}
We present a method for calculating  \it ab initio \rm interatomic forces
 which scales quadratically  with the size of the system and provides a 
physically transparent representation of the force in terms of the spatial variation of the electronic
 charge density.  The method is based on a reciprocity theorem for evaluating
 an  effective potential acting on a charged ion in the core of each  atom.  We illustrate 
 the method with calculations  for diatomic molecules.
\end{abstract}  
]
  Presently,  first-principles computational methods can be used to study the equilibrium ground-state structures and transient excited-state relaxation pathways even for relatively complex systems  containing as many as several hundred atoms 
  \cite {rmp}.
  In these calculations one is guided through
 a   large phase space of possible atomic configurations by following the gradients with 
respect to the classical nuclear coordinates  of
 the total energy of a  system of interacting electrons and ions.  On the Born-Oppenheimer ground-state surface these gradients can be obtained by exploiting a force  theorem (the Hellmann-Feynman
 theorem)
 \cite{feyn}  which states that for any linear variation of a control parameter $\lambda$ in the
 quantum mechanical Hamiltonian
 \begin{equation}
 \frac {\partial U} {\partial \lambda} = \left <  \frac {\partial H} {\partial \lambda} \right >
\end{equation}
where $U$ is the total energy of the system, $H$ is the Hamiltonian and the brackets denote an expectation value in the
electronic ground state. Using the control parameter $\lambda$ to denote the nuclear coordinate
$R_{I \alpha}$ the theorem immediately provides the force acting on the $I$-th lattice site  with
 the $\alpha$-th polarization.
 
 In this paper we discuss a reciprocity theorem which provides a particularly efficient
 scheme for evaluating the forces in equation (1). 
 One often regards (1) as measuring a response of the quantum electronic charge
distribution to an ionic displacement, and indeed the expectation value requires an average of a nuclear
deformation potential in the electronic ground state. Alternatively, the left hand
 side of equation (1) is a force on a classical ion responding to the total electric
field seen at the ion site.  One can exploit this latter
 point of view to greatly simplify the calculation and interpretation of this force. 
   We present a reciprocity theorem for inverting this problem which 
applies even in  the general case  
   where the interaction between the electronic and nuclear degrees of freedom is
 described by effective potentials \cite{rmp}.
  Using this theorem the evaluation of (1) for a system containing $N_a$ atoms scales \cite{ks} as  $N_a^2$ 
rather than as $N_a^3$ as in current electronic structure calculations \cite{rmp}. The method also greatly aids the
 interpretation of these forces by 
directly relating them to the spatial distribution of the ground-state charge density. 

 Our starting point is a Hamiltonian \cite {dft} describing an interacting system of ions and electrons, 
with the ionic degrees of freedom treated classically, $H= H_k + H_{\rm ee} + H_{\rm ion-ion} + H_{\rm el-ion}$:
\begin{eqnarray}
H &=& \sum_i - \frac{\hbar^2}{2m} \nabla _i ^2 + \frac {1}{2} \sum_{i \ne j} \frac {e^2}{|r_i - r_j|} \nonumber\\
 &+& \frac{1}{2} \sum_{I \ne J} \frac { Z_I Z_J e^2}{|R_I - R_J|} +  V_{ \rm el-ion} 
\end{eqnarray}
 The potential $V_{\rm el-ion}$ may be either a Coulomb potential or a more sophisticated effective potential for treating only the valence electronic
 degrees of freedom. In either case the Hamiltonian for the system depends parametrically on
 the ionic coordinates $R_I$ only through the last two terms on the right hand side of equation (2) so that  by using the force theorem, one has:
\begin{equation}
F_{I \alpha} = - \nabla_{R_{I \alpha}} U =   - \frac {\partial H_{\rm ion-ion}}{\partial R_{I \alpha}}
 -  \left <
\frac {\partial V_{{ \rm el-ion}}} {R_{I \alpha}} \right > 
\end{equation}
 If we assume that the electrons and ions interact through the simple Coulomb potential, then  one can rewrite equation (3) from the point of view of the classical
ions by first isolating the part of the total energy which depends on the nuclear
coordinates: 
\begin{equation}
  \frac{1}{2} \int \rho_{\rm ion} (r) V_{\rm ion} (r) d^3r + 
 \int \rho _{\rm el} (r) V_{\rm ion} (r) d^3r  
\end{equation}
where $\rho_{\rm el}$ is the electronic charge density, $\rho_{\rm ion} (r) = \sum_I Z_Ie \delta(r-R_I)$ denotes
the ionic charge density, and $V_{\rm ion}(r)$ is the electrostatic potential produced by the
ionic source distribution $\rho_ {\rm ion}$.  
The second term can be rewritten by observing that the
 electronic Hartree potential satisfies $-  \frac {1}{4 \pi}  \nabla^2 V_H = \rho_{\rm el} (r)$, so  that after integrating twice by parts, one has 
\begin{equation}
U_{{ \rm el-ion}} = \int V_H(r) ( - \frac{1}{4 \pi} \nabla^2 V_{\rm ion}(r) ) d^3 r
\end{equation}
For a system of point ions, the second factor in the integrand is  $\sum_I  Z_I e \delta (r - R_I)$.
 Thus differentiation with respect to the ionic coordinates explicitly
shows that the force  on each ion is a response to the electric field, $-\nabla V_H(r)$ acting at the ionic site as one expects. 

The first  term in expression (4) also contributes to the electric field acting
on the $I$-th site through the gradient $-\nabla \sum_{J \ne I} V_{\rm ion} (r - R_J)$ i.e.  the gradient of the potential produced by the
unscreened ions excluding the contribution from the  {\it I}-th site.  Combining these two contributions
for the {\it I}-th ion, one has
\begin{equation}
V_I(r) = \sum_{J \ne I} V_{\rm ion} (r - R_J)   + V_H(r)
\end{equation}
so that $F_{I \alpha} = - Z_ie\nabla _{I \alpha} V_I$
 
The restriction on the sum in equation (6) represents an awkward computational constraint
 since in principle the sum is different for each ionic site in the system. However, this
 can be dealt with efficiently  by redistributing each ionic point charge  uniformly
 on a spherical shell  at some desired atomic sphere radius, $a_c$ in the first term of (6).   This shell provides a potential $V_s$ which has the correct 1/r dependence for $r > a_c$
and  is constant for $r < a_c$, so that the spatially varying part of $V_s$ near the i-th site
 correctly represents the constrained sum in (6).   With this replacement  
 the effective potential $V_I$  which governs relaxation of the classical ions can
 be obtained for  all the lattice sites {\it I}  in a single calculation.

We illustrate  the method with a calculation on the diatomic molecule ${ \rm H_2}$.  In figure 1 we present a map
 of the equipotentials of equation (6)  within an  atomic sphere centered around  one of the H ions in the
 molecule. The three panels give the equipotentials calculated  using the local density approximation (LDA)  to the exchange and correlation  energies,  for three different interatomic separations.   
 In all cases
 a minimum in this potential lies near the center of  the atomic sphere. For the compressed  ${ \rm H_2}$ the
 minimum lies at larger separation, and for the expanded molecule at a smaller
 separation.  The force, calculated from the gradient of this potential, is plotted 
 as a function of interatomic separation in figure 2, where we overlay the forces
 obtained by direct  use of the force theorem (1), and as expected the two agree  exactly. 
  The inner panel  gives the offset  between the interatomic separation
 and the separation which minimizes the effective potential computed for each separation.
  This shows quite accurately the equilibration of the molecule at the LDA bond
 length of 0.78A. 
 
  The essence of the force theorem is that to lowest
 order in the
 nuclear displacements, the electronic charge density can be regarded as rigid.
In (1) one then  performs an average of the deformation potential in this rigid
density. In (5) we use this
 rigid charge density as a source term in the Poisson equation to extract the effective
 potential near the ionic site. The equivalence between these two reciprocal points of view  
 applies to all types of
 interatomic interactions, including metallic, covalent, and even van der Waals
bonding,  as long as an accurate
 representation of the charge density is in hand.

	This method can be generalized to a system of electrons and ions interacting
 through any local pseudopotential $V_{{ \rm ps}}(r)$.  A pseudopotential generally does not
 retain the 1/r singularity as r $\rightarrow$ 0, and thus the partial integrations
 leading to (5)  no longer project the Hartree potential exactly onto a lattice site.  However, by
 replacing $V_{\rm ion}$ by $V_{{ \rm ps}}$ in (5) 
one sees  that the interaction energy can always be obtained by
integrating the Hartree potential over an effective ionic charge density which is calculated 
 by taking the Laplacian of $V_{{ \rm ps}}$.  Alternatively we observe that 
 the interaction energy $U_{{ \rm ps}}$ has the form:
\begin{equation}
U_{{ \rm el-ion}} =  \sum_I \int \rho _{\rm el}(r) V_{{ \rm ps}} (r - r') \delta (r' - R_I) d^3r d^3r'
\end{equation}
so that by integrating over the coordinate r \it first, \rm one has
\begin{equation}
U_{{ \rm el-ion}} = \sum_I \int V_{{\rm eff}} (r') Z_I e \delta (r' - R_I) d^3r'
\end{equation}
Equation  (8) explicitly shows that after the spatial average, a  point nucleus
 experiences an effective potential where $\rho_{\rm el}$ is the source  term, 
and for which the interaction kernel is the effective pseudopotential. If the
 pseudopotential is replaced by the bare Coulomb potential, the effective potential
 acting on a lattice site is just the electronic Hartree potential as we found earlier.

	In figure 3 we apply this method  to  molecular ${\rm H_2}$,  but now
 calculated replacing the H ions by pseudo-ions. In the main part of the figure we
 overlay the forces calculated from the force theorem and calculated using the reciprocity
 relation, and again the two agree exactly. The inset gives 
 the distance from the center of the sphere to the potential minimum as a function of
 interatomic  separation, and  confirms that this offset crosses zero at the expected equilibrium
 separation.

Accurate first-principles pseudopotentials  frequently require a nonlocal
representation $V_{{\rm ps}} (r,r')$, so that the interaction energy analogous to
equation (7)  has  the form:
 \begin{eqnarray}
U_{{\rm el-ion}} = \sum_{n,I}
   \int \psi_n^*(r) &V&_{{\rm ps}} (r-R_I,r'-R_I)  \nonumber\\
  &&    \psi_n (r')  d^3r d^3 r'
\end{eqnarray}
and the sum is over occupied single-particle states $ \psi_n$. Thus the interaction energy involves 
an integral over the full  one-particle density matrix $\rho(r,r')$ rather than
simply the charge density alone, i.e. 
\begin{equation}
U_{{\rm el-ion}} = {\rm Tr} \rho(r,r') V_{{\rm ps}}(r,r')
\end{equation}
Inverting this relation along the lines of equations (5) and (7) requires
the solution of a generalized Poisson equation with a source term
 which is the one-particle density matrix. Instead, by 
integrating equation (9)  over a single intermediate coordinate $s$  we obtain
\begin{eqnarray}
U_{{\rm el-ion}} = \sum_{n,I}  \int & \psi_n^*(r)& V_{\rm ps}(r-s,r'-s) \psi_n (r') \nonumber\\
 &&   d^3r d^3r' \delta(s-R_I) d^3s \nonumber\\
 &=& \sum_I \int V_{\rm eff} (s) Z_I  e \delta (s - R_I) d^3s
\end{eqnarray}
with
\begin{equation}
V_{{\rm eff}} (s) = \frac{1}{Z_I e} \int \rho (r,r') V_{{\rm ps}} (r-s,r'-s) d^3r d^3r'
\end{equation}
This potential simplifies still further if the effective nonlocal potential
is expressed as a sum of separable potentials as in Kleinman and Bylander's
construction \cite{kb}
\begin{equation}
V_{{\rm el-ion}}(r,r') = \sum_{I,c}  \phi_c(r-R_I) \alpha_c \phi_c^*(r'-R_I)
\end{equation}
where the $\phi_c(r-R_I)$ are a set of projection functions centered on the $I$-th
 ionic site, and $\alpha_c$
are the associated weights.
Expressing the one-electron states with  the Fourier expansion
\begin{equation}
\psi_{nk} (r) = \sum_G e^{ i(k+G) \cdot r} c_{nk} (G)
\end{equation}
the nonlocal  contribution to the interaction energy is
\begin{eqnarray}
 U_{{\rm el-ion}}   &=& \sum_{G G'} \sum_{nk} \sum_{I,c}  e^{-i(G-G') \cdot R_I} \nonumber\\
	&& \langle k+G | \phi_c  \rangle \alpha_c \langle \phi_c | k+G' \rangle c^*_{nk}(G) c_{nk}(G')
\end{eqnarray}
In equation (15)  the electronic ground state enters through the density matrix
 elements $\sum_{nk} c^*_{nk} (G) c_{nk}(G')$, while the plane-wave matrix 
 elements of the nonlocal potential appearing in the sum are normally tabulated in an  atomic
  pseudopotential
 code.  Thus the generalized kernel which connects the one-particle
 density matrix to the nuclear coordinate is  the nonlocal potential. 

	As an illustration  of the method we have applied the prescription in equation (15) to
 the ${\rm Cl_2}$ dimer. Here the Cl ions are represented by nonlocal pseudopotentials \cite{optpsp} in
 the Kleinman-Bylander form. In figure (4) we display the force on each ion as a function
 of the interatomic separation, overlaying the forces computed from the Hellmann-Feynman force  theorem and from  
 the reciprocity relation.  As expected,  the two agree exactly, confirming 
 the validity of the construction.
 
 For a system containing $ N_a $ sites, the evaluation of $U_{{\rm el-ion}}$ (and its
 gradients) using the reciprocity theorem scales \cite{caveat} as $ N_a^2 \log_2 N_a $ rather than as $ N_a^3$ as would be required
 for a  direct evaluation of equation (15). 
 This occurs because the Kleinman-Bylander projection functions $\phi_c (r-R_I)$
for different atoms $I$, but the same atom type, differ only in the  location of the guiding centers, $R_I$.
 Then one can rewrite equation (15)  as a single sum 
\begin{equation}
U_{{\rm el-ion}} = \sum_{\Delta G} \sum_c e^{- i \Delta G \cdot R_i} \Omega (\Delta G)
\end{equation}
 where the kernel $\Omega$ requires the convolution of $\phi_c(G) c_{nk} (G)$ with its
 complex conjugate, a computation which can be completed in ${\cal O} ( N_a \log_2 N_a)$ steps using the fast Fourier transform. 
 Thus the evaluation of the energy and its  gradients for all sites
 can be accomplished in $ {\cal O} ( N_a^2 \log_2 N_a )$ steps.  This is a significant advance since it
 leaves the wavefunction orthogonalization operation as the only remaining $ N_a^3$ scaling
 operation in  plane wave based electronic  structure calculations \cite{rmp}. 

 However, in addition to its computational efficiency, a significant advantage of the
 reciprocity rule is that it relates the forces seen by the ions to
 the spatial distribution of the electronic charge density in a direct
 and physically transparent way. This is likely to 
 be extremely useful for interpreting atomic relaxations near surfaces and
 defects in solids, for which the relaxations can be quite complicated, but
 which 
 are nonetheless determined  by the spatial  redistribution of the valence electron density.
 In principle one might be able to exploit the composition of the nonlocal
 potential in equation (15) to clearly isolate contributions  to the force due to  the {\it s-} {\it p-} and
 {\it d}-like components of the charge density.  

	In summary, by exploiting a reciprocity relation for the electron-ion interaction
 one may reformulate the problem of calculating atomic forces in electronic structure codes
 into  a single computation which will simultaneously yield the force distribution at all 
 lattice sites in the system.  The method scales efficiently with system size and directly
 relates the computed forces to the spatial distribution of the electronic charge density.

	We would like to thank N.J. Ramer for his assistance in the construction of the Cl
 pseudopotentials, and K.M. Rabe for a useful correspondence. This work was supported in
 part by the NSF under grant number 93 13047.

\begin{figure}
\epsfxsize=3in
\centerline{\epsffile{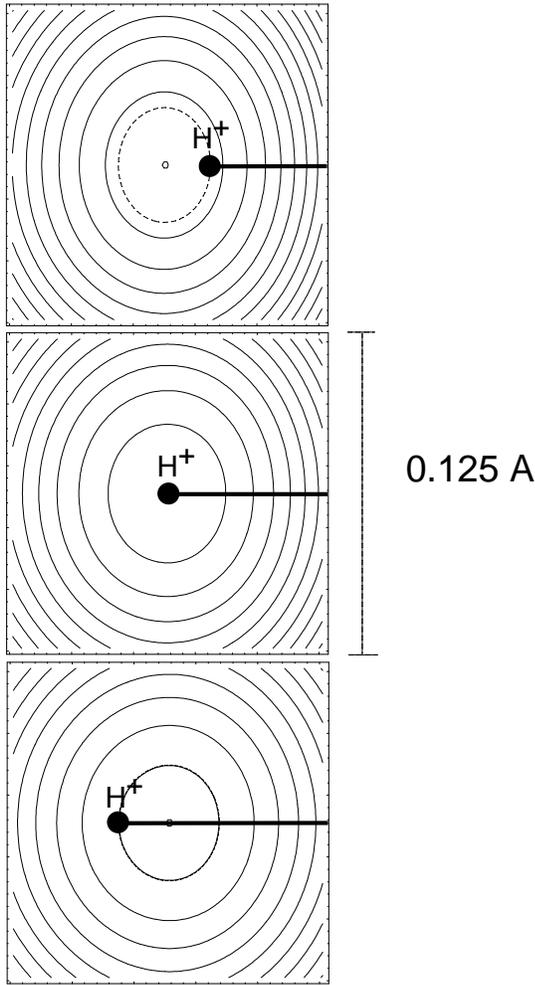}}
\caption{Contour plots of the effective potential of equation (6) seen by
an ${\rm H^+}$ ion in ${\rm H_2}$ calculated at bond lengths of 0.70A (top), 0.78A (equilibrium,
middle) and 0.90A (bottom). The equipotentials are equally spaced at intervals
 of 50 meV. The location of the ${\rm H^+}$ ion is given by the bold dot. The string running
 off each panel to the right terminates on the other   ${\rm H^+}$ ion.}
\end{figure}
 
\begin{figure}
\epsfxsize=3in
\centerline{\epsffile{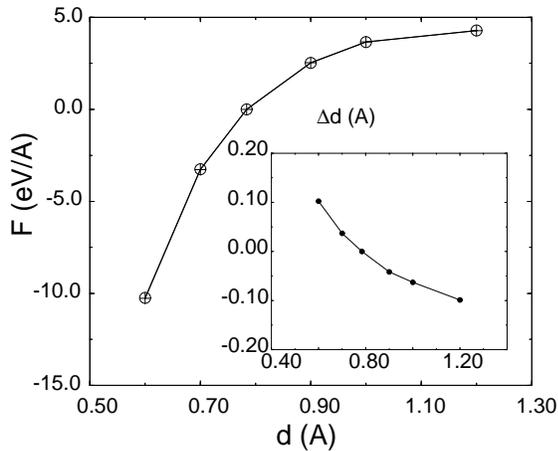}}
\caption{ The force on an ${\rm H^+}$ ion in  ${\rm H_2}$ calculated using the full
 Coulomb potential plotted as a function of interatomic separation. The plot
 overlays the results from the Hellmann-Feynman force theorem (open circles) and
 the force obtained from the reciprocity relation discussed in the text (crosses). Here a
 positive force represents an attractive force. The
 vertical axis of the inner panel  gives the offset between the interatomic separation and the distance
 between potential minima calculated for each interatomic separation.}
\end{figure}
 
\begin{figure}
\epsfxsize=3in
\centerline{\epsffile{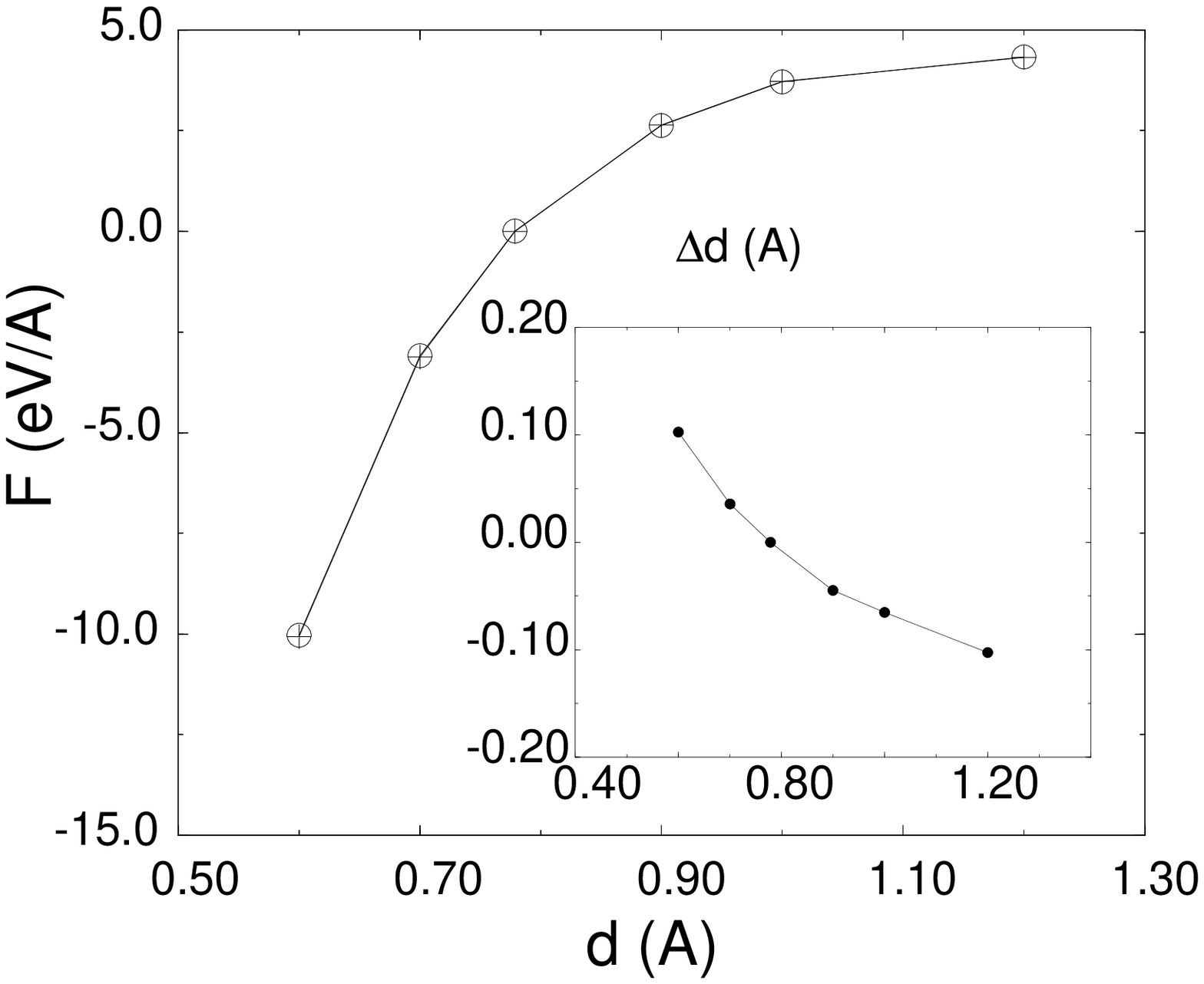}}
\caption{ The force on an  ${\rm H^+}$ ion in   ${\rm H_2}$ calculated using a local
 pseudopotential plotted as a function of interatomic separation. The plotting conventions
 are the same as those of figure (2).}
\end{figure}

\begin{figure}
\epsfxsize=3in
\centerline{\epsffile{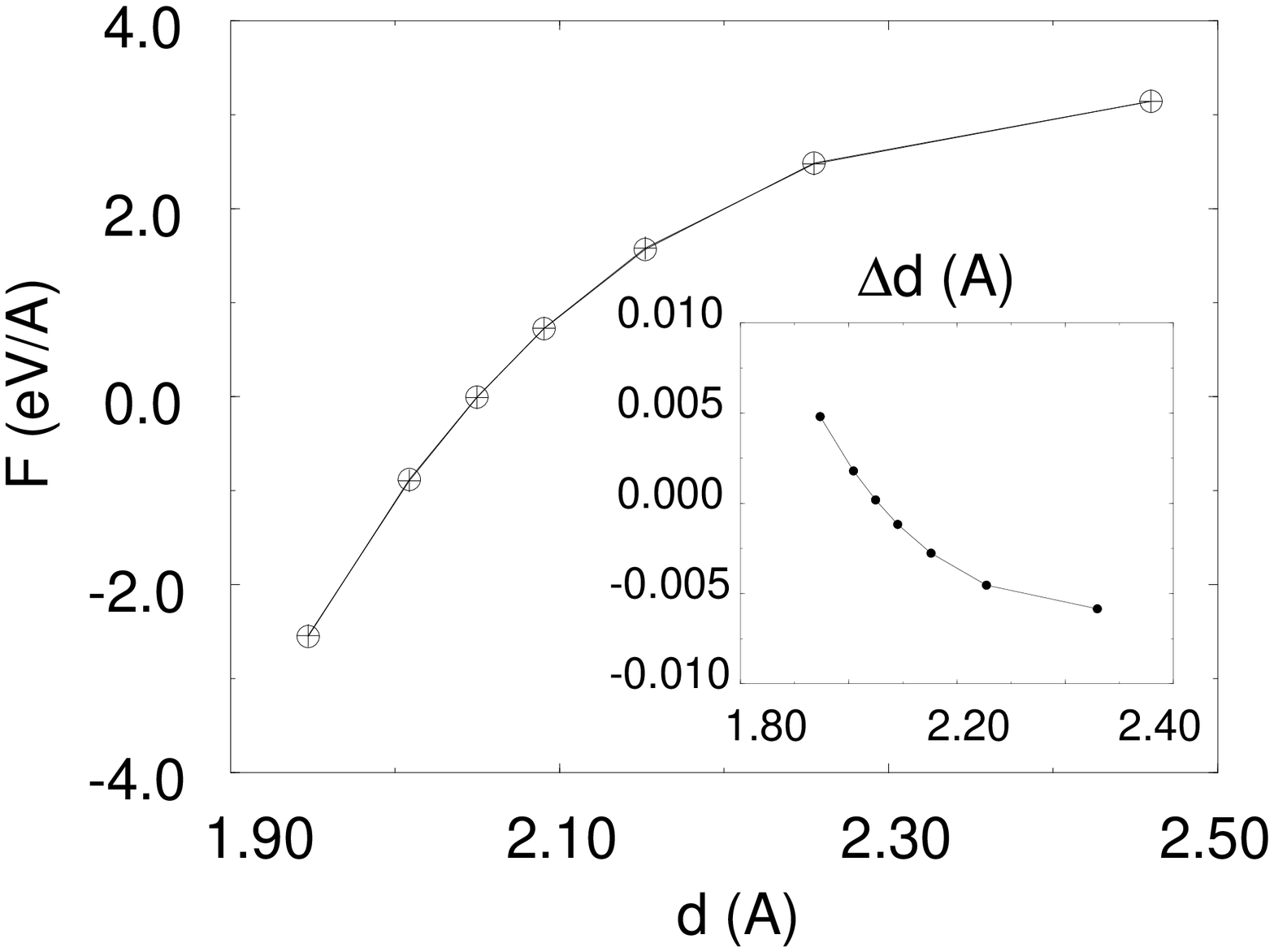}}
\caption{ The force on a $\rm  Cl$ ion in ${\rm Cl_2}$ calculated using a nonlocal
 pseudopotential plotted as a function of interatomic separation. The plotting conventions
 are the same as those of figure (2).}
\end{figure}


\begin{references}
\bibitem{dft} More generally the  method can also be applied to any energy functional in which the
 ion coordinates are treated as classical degrees of freedom.
\bibitem{caveat}  This also differs from  the ${\cal O}( N_a^2)$  method of
 reference [3] which is carried out in real space; our algorithm is  carried out completely in
reciprocal space, and is especially well suited for plane wave based electronic structure methods.
\end{references}
\end{document}